\documentclass[doublecol]{epl2} 

\title{Drainage in a model stratified porous medium}
\shorttitle{Drainage in a model stratified porous medium} 

\author{Sujit S. Datta and David A. Weitz}
\shortauthor{S. S. Datta and D. A. Weitz}

\institute{                    
  \inst{1} Department of Physics, Harvard University, Cambridge MA 02138, USA}

\abstract{
We show that when a non-wetting fluid drains a stratified porous medium at sufficiently small capillary numbers $Ca$, it flows only through the coarsest stratum of the medium; by contrast, above a threshold $Ca$, the non-wetting fluid is also forced laterally, into part of the adjacent, finer strata. The spatial extent of this partial invasion increases with $Ca$. We quantitatively understand this behavior by balancing the stratum-scale viscous pressure driving the flow with the capillary pressure required to invade individual pores. Because geological formations are frequently stratified, we anticipate that our results will be relevant to a number of important applications, including understanding oil migration, preventing groundwater contamination, and sub-surface CO$_2$ storage. }

\begin{document}

\maketitle

\section{Introduction}
Drainage, the displacement of a wetting fluid from a porous medium by an immiscible non-wetting fluid, arises in many technological problems, including oil migration and recovery \cite{oil1, oil2}, waste CO$_{2}$ sequestration \cite{c1,c2,c3,c4}, and groundwater contamination \cite{waste1,waste2,culligan}. The ability to accurately predict the flow behavior of the non-wetting fluid is critically important in all of these examples. To displace the wetting fluid from a pore, a threshold capillary pressure must build up in the non-wetting fluid at the pore entrance; this pressure is given by $P_{\gamma}\sim\gamma/a_{t}$, where $\gamma$ is the interfacial tension between the fluids and $a_{t}$ is the radius of the pore entrance \cite{amber,toledo,mason}. For a homogeneous porous medium, characterized by pores of a single average size, $P_{\gamma}$ is typically much larger than the viscous pressure associated with flow into a pore. Consequently, the flow path taken during drainage depends primarily on the slight pore-scale variations of $a_{t}$ \cite{invasion1,invasion2,robbins1,invpercvisc,lei}. However, many porous media are stratified, consisting of parallel strata characterized by different average pore sizes \cite{layered1, layered2,gasda}. Such additional variation in the pore structure, on scales much larger than a single pore, may strongly modify the flow behavior \cite{firoozabadi, keuper,pinder,shokri,huppert,illa,bodiguel,invpercvisc,newref1,newref2}. Despite its enormous practical importance, a clear picture of how the subtle interplay between capillary and viscous forces determines the flow through a three-dimensional (3D) stratified porous medium remains elusive. This requires direct visualization of the multiphase flow, both at the scale of the individual pores and the overall strata. Unfortunately, the medium opacity typically precludes such visualization. As a result, knowledge of how exactly drainage proceeds within a stratified porous medium is missing.

Here, we use confocal microscopy to investigate drainage within a 3D porous medium having parallel strata oriented along the flow direction; this allows us to directly visualize the multiphase flow at pore-scale resolution. We find that for sufficiently small capillary numbers, $Ca$, the non-wetting fluid flows only through the coarsest stratum of the medium. By contrast, above a threshold $Ca$, the non-wetting fluid is also forced laterally into part of the adjacent, finer strata. By balancing the viscous pressure driving the flow with the capillary pressure required to invade a pore, we show how both the threshold $Ca$ and the spatial extent of the invasion depend on the pore sizes, cross-sectional areas, lengths, and relative positions of the strata. Our results thus help elucidate how the path taken by the non-wetting fluid is altered by stratification in a 3D porous medium.

\section{Experimental Methodology}
We prepare rigid 3D porous media by lightly sintering densely-packed hydrophilic glass beads, with polydispersity $\approx4\%$, in thin-walled rectangular quartz capillaries; these have cross-sectional areas $A=1, 3,$ or $4$ mm$^{2}$. The packings have porosity $\phi\approx0.41$, as measured with confocal microscopy \cite{amber}. To create stratified porous media, we arrange the beads into parallel strata characterized by different bead sizes and the same length $L$; the interface between the strata runs along the direction of fluid flow. To enable imaging of the multiphase flow within the media, we use fluids whose compositions are carefully chosen to match their refractive indices with that of the glass beads; the wetting fluid is a mixture of 91.4 wt\% dimethyl sulfoxide and 8.6 wt\% water, while the non-wetting oil is a mixture of aromatic and aliphatic hydrocarbons. The viscosities of the wetting fluid and the oil are $\mu_{w}=2.7$ mPa.s and $\mu_{o}=16.8$ mPa.s, respectively. The interfacial tension between the two fluids is $\gamma=13.0$ mN/m, as measured using a du No\"uy ring. We use confocal microscopy to estimate the three-phase contact angle between the wetting fluid and glass in the presence of the oil, $\theta\approx5^{\circ}$ \cite{amber}. 

\begin{figure}
\begin{center}
\includegraphics[width=3.4in]{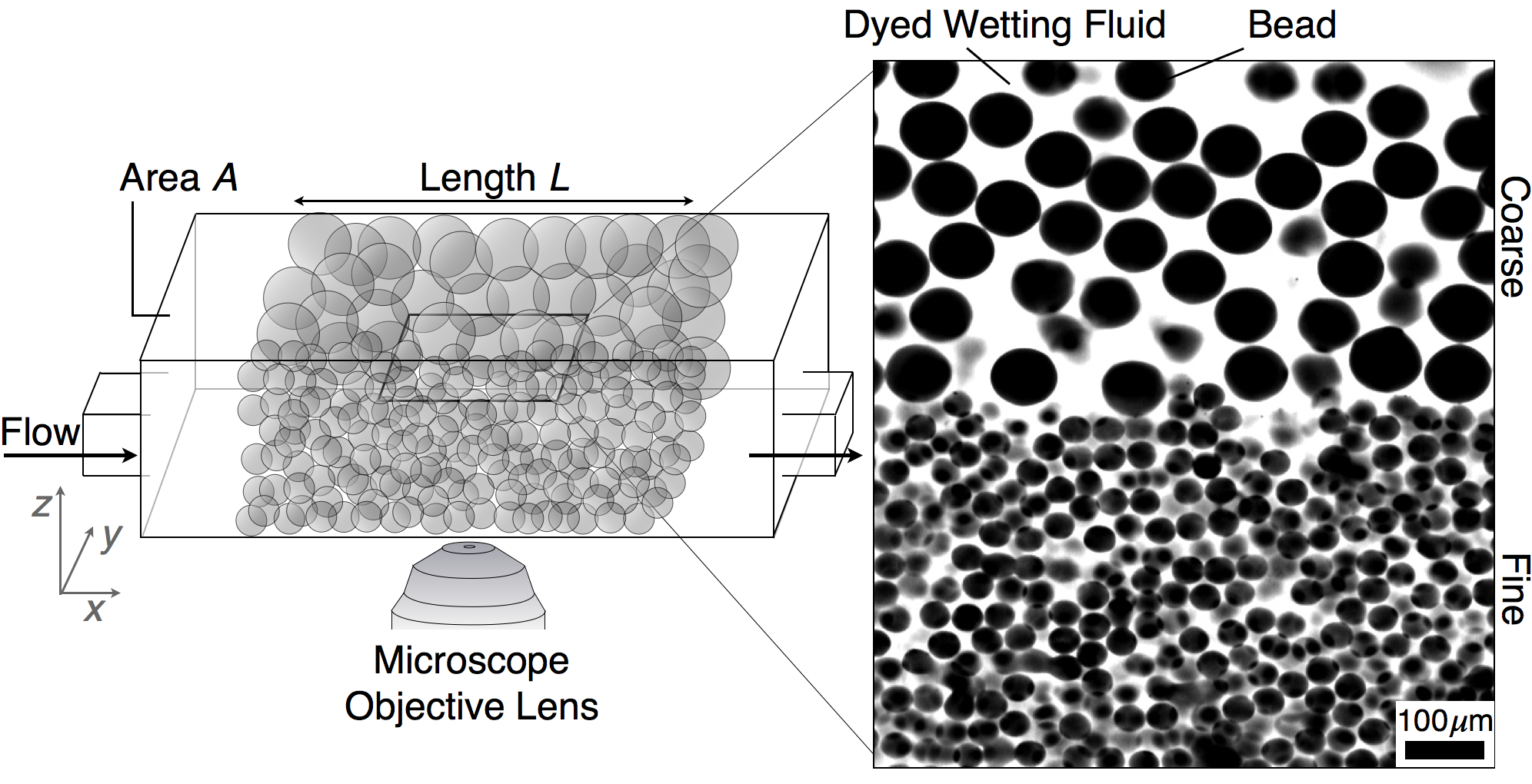}
\caption{(Left) Schematic of a stratified porous medium, with parallel strata comprised of differently-sized beads. (Right) Optical section acquired within a porous medium with a coarse and a fine stratum; the pore space is saturated with the fluorescently-dyed wetting fluid, and the black circles show cross sections of the beads making up the medium.}  
\end{center}
\end{figure}

Prior to each experiment, the porous medium is evacuated under vacuum and saturated with CO$_{2}$ gas; this gas is soluble in the wetting fluid and prevents the formation of any trapped bubbles. We then saturate the medium with the wetting fluid; to visualize the fluid, we dye it with fluorescein. We visualize the pore structure in 3D using a confocal microscope, acquiring a 3D stack of 39 11 $\mu$m-thick optical slices, each spanning a lateral area of 912 $\mu$m $\times$ 912 $\mu$m, spaced by 6 $\mu$m along the $z$-direction, within the porous medium, away from its boundaries [schematized in Fig. 1]. We identify the glass beads by their contrast with the dyed wetting fluid; an example slice is shown in Fig. 1 \cite{amber}. Moreover, we acquire stacks at multiple locations along the length of the medium; by combining these stacks, we obtain a map of the pore structure of the entire medium.

To investigate the drainage process, we subsequently flow the oil at a constant volumetric rate $Q$, re-acquiring 3D image stacks at the same positions within, and along the length of, the medium. The oil is undyed; we thus identify it by its additional contrast with the dyed wetting fluid in the measured pore volume. This enables us to directly visualize the multiphase flow, both at the scale of the individual pores and the overall strata.
\begin{figure}
\begin{center}
\includegraphics[width=3.4in]{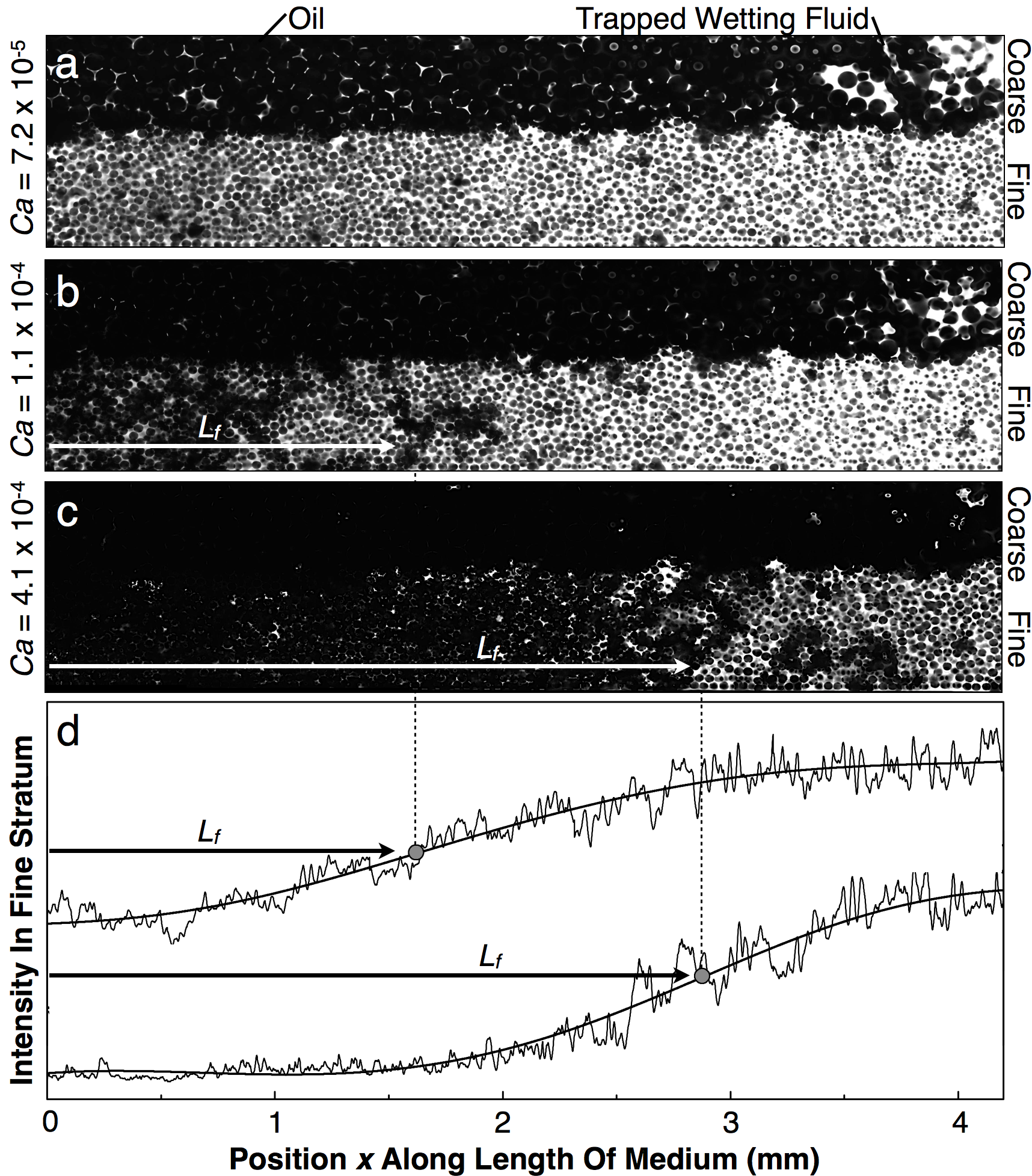}
\caption{Optical sections through an entire porous medium with a coarse and a fine stratum, obtained after flow has reached an unchanging steady state; imposed flow direction is from left to right. (a) Oil flows exclusively through the coarse stratum for $Ca=7.2\times10^{-5}$; (b-c) oil also invades part of the fine stratum, up to a length $L_{f}$ from the inlet, for $Ca=1.1\times10^{-4}$ and $4.1\times10^{-4}$, respectively. The pore space is initially saturated with the fluorescently-dyed wetting fluid; black circles are the beads comprising the medium, while additional black fluid is the invading oil. Inlet is at position $x=0$.(d) Fluorescence intensity in the fine stratum, integrated along $y$ and $z$, as it varies with $x$ for $Ca=1.1\times10^{-4}$ and $4.1\times10^{-4}$ as shown by the top and bottom traces, respectively. Smooth curves show filtered data, and arrows show distance invaded by the oil $L_{f}$, determined from the inflection points of the curves. The two traces, and corresponding curves, are vertically shifted for clarity.}  
\end{center}
\end{figure}

\section{Results and Discussion}
The effect of stratification on the flow behavior is exemplified by draining a porous medium with a coarse and a fine stratum, comprised of beads with radius $a_{c}=38 ~\mu$m and $a_{f}=19 ~\mu$m, respectively [Fig. 1]. The medium has length $L=4.2$ mm and a total cross-sectional area $A=1  $ mm$^{2}$; the coarse stratum occupies an area $A_{c}=0.44A$. We flow the oil at a constant rate $Q=0.2$ mL/hr, corresponding to a capillary number $Ca\equiv\mu_{o} (Q/A)/\gamma=7.2\times10^{-5}$ \cite{capnum}. The oil invades the medium through a series of abrupt bursts into the pores, indicating that a threshold capillary pressure must build up in the oil before it can invade a pore \cite{amber}. Interestingly, the oil flows {\it exclusively} through the coarse stratum, as shown by the optical slice in Fig. 2(a), over an observation time of 30 min \cite{time}.

\begin{figure}
\begin{center}
\includegraphics[width=2.7in]{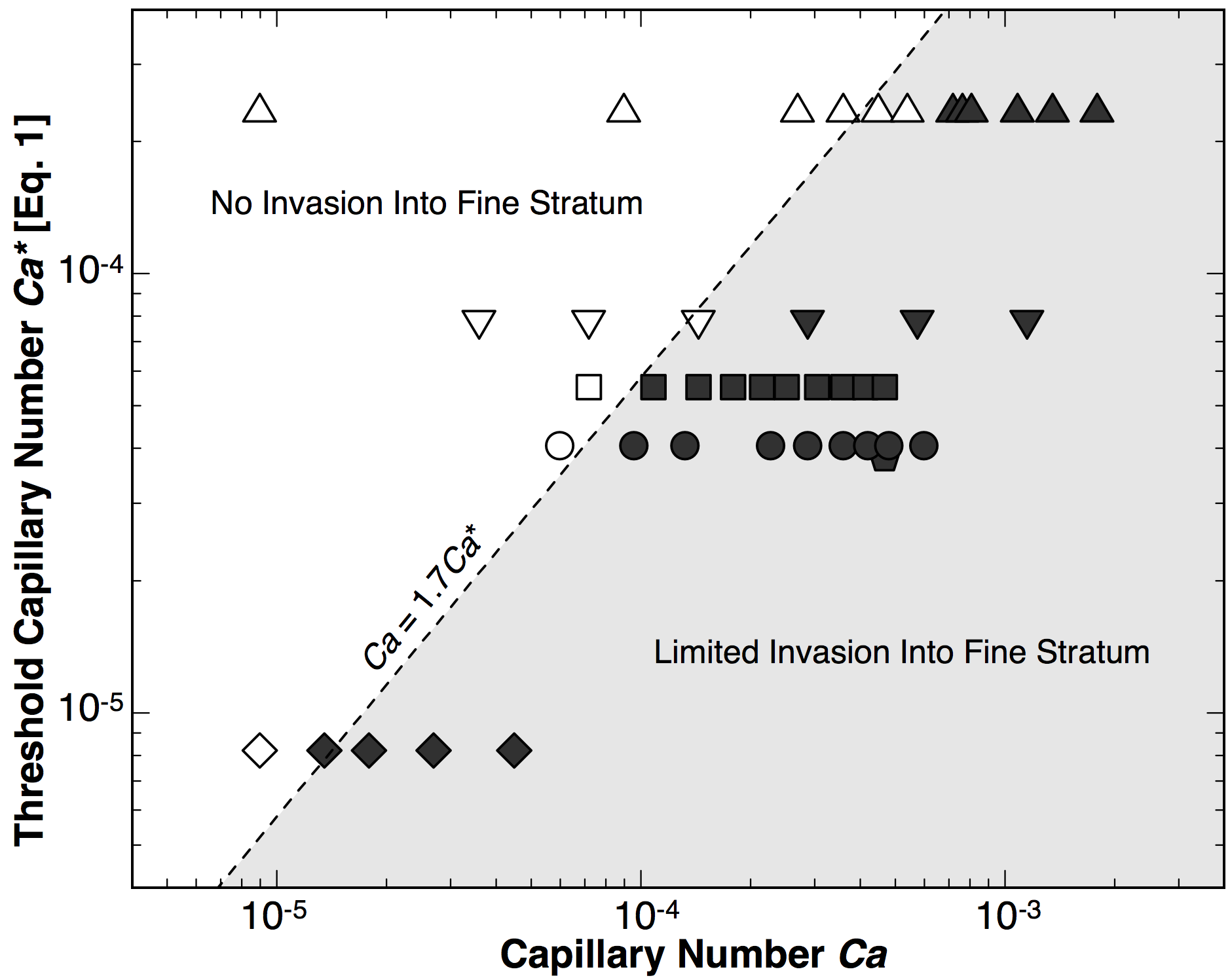}
\caption{For sufficiently small $Ca$, oil does not invade into fine stratum (open symbols), while for sufficiently large $Ca$, oil partially invades the fine stratum as well (filled symbols). Data are shown for different porous media, with different cross-sectional areas, lengths, and bead sizes; these are thus characterized by different values of $Ca^{*}$, defined in Eq. (1). The threshold for invasion into the fine stratum is approximately $Ca=1.7Ca^{*}$ (dashed line).}  
\end{center}
\end{figure}

\begin{figure}
\begin{center}
\includegraphics[width=3.4in]{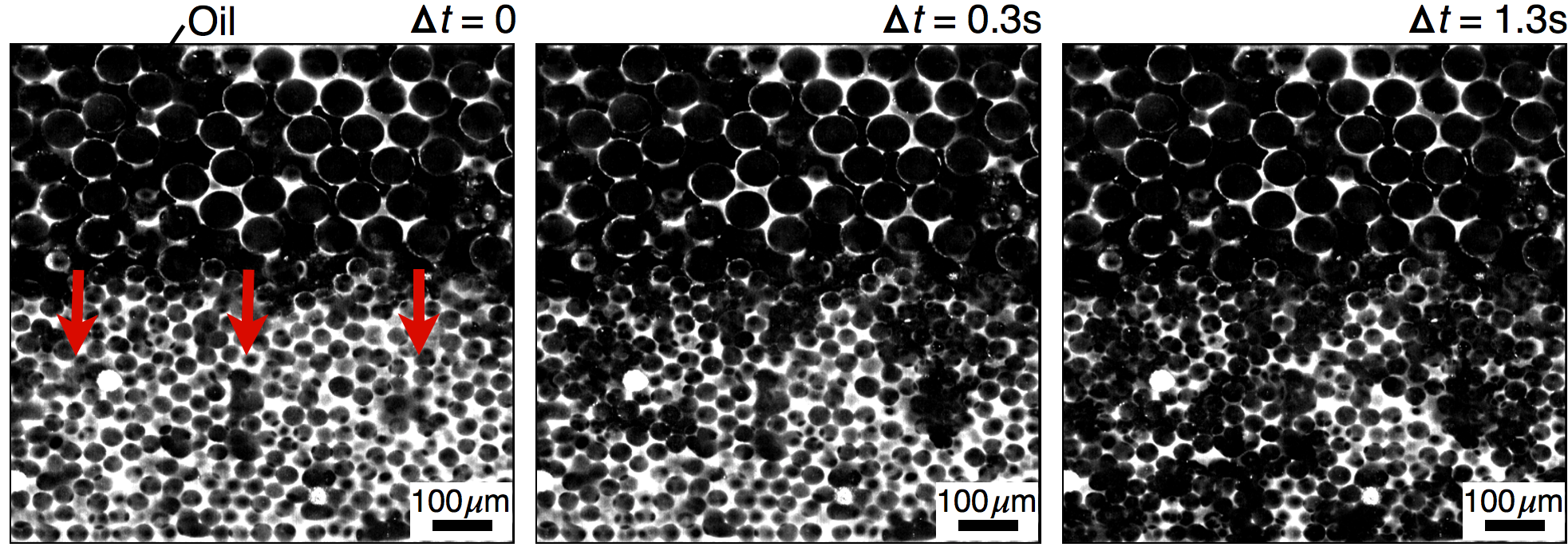}
\caption{Time sequence of confocal micrographs, taken at $Ca>Ca^{*}$, showing flow of oil (black) from the coarse stratum laterally into the fine stratum, indicated by arrows in first panel; $Ca=4.7\times10^{-4}$ and $Ca^{*}=3.8\times10^{-5}$. Imposed flow direction is from left to right.}  
\end{center}
\end{figure}

\begin{figure}
\begin{center}
\includegraphics[width=2.7in]{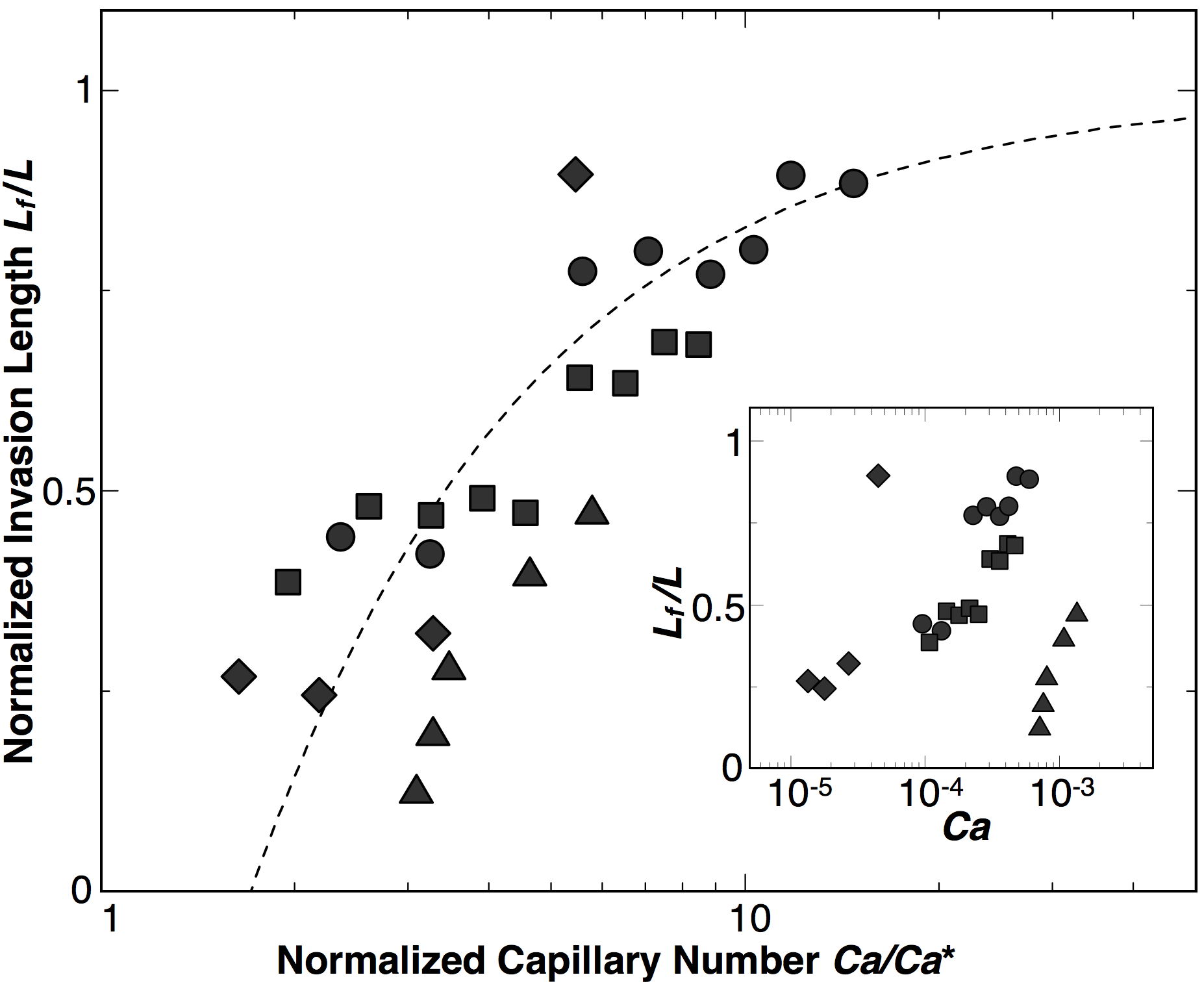}
\caption{(Inset) Distance invaded into the fine stratum, $L_{f}$, increases with increasing $Ca$; different symbols represent different media characterized by different values of $Ca^{*}$. Symbols are the same as in Figure 3. $L$ is the length of the medium. (Main panel) Data collapse when $Ca$ is rescaled by $Ca^{*}$, and agree well with the theoretical prediction $L_{f}/L=1-1.7Ca^{*}/Ca$ (dashed line).}  
\end{center}
\end{figure}
To further explore the drainage process, we increase the oil flow rate to $Q=0.3$mL/hr, corresponding to $Ca=1.1\times10^{-4}$. Surprisingly, in contrast to the $Ca=7.2\times10^{-5}$ case, the oil invades the fine stratum, although only partially, over a limited distance, $L_{f}$, from the inlet, as shown in Fig. 2(b). After an unchanging steady state is reached, we flow the oil at even higher flow rates and probe the resulting steady state invasion patterns. Interestingly, the fine stratum remains only partially invaded for the entire range of $Ca$ explored; however, we find that $L_{f}$ increases with increasing $Ca$, as shown in Fig. 2(b-c). This observation contradicts the idea that the fine stratum is completely impervious to the oil \cite{epa}. 

To quantify the partial oil invasion into the fine stratum, we integrate the fluorescence intensity in the fine stratum along both the $y$ and $z$-directions, for each position $x$. Two examples, corresponding to the invasion patterns shown in Fig. 2(b) and (c), are shown by the upper and lower traces in Fig. 2(d), respectively. To determine the distance invaded by the oil into the fine stratum, $L_{f}$, we apply a low-pass filter to these data \cite{interface} and determine the distance from the inlet to the inflection point of each filtered curve [points in Fig. 2(d)]. Consistent with the optical slices shown in Fig. 2(b-c), we find that $L_{f}$ increases steadily with increasing $Ca$, as shown in the inset to Fig. 5. 

To understand this complex flow behavior, we analyze the distribution of pressures in the oil as it displaces the wetting fluid. For the oil to invade a pore formed by beads of radius $a$, the capillary pressure at the oil-wetting fluid interface must exceed $P_{\gamma}=2\gamma\cos\theta/a_{t}$, where $a_{t}\approx0.19a$ \cite{amber,toledo,mason,bryant,meheust}. The capillary pressure required to invade a pore of the coarse stratum, $P_{\gamma, c}\sim1/a_{c}$, is thus smaller than that required to invade a pore of the fine stratum, $P_{\gamma, f}\sim1/a_{f}$; consequently, we expect the coarse stratum to be drained first. This expectation is in direct agreement with the observed drainage behavior, shown in Fig. 2(a).

An additional viscous pressure, $P_{v}$, drives the continued flow of oil through the coarse stratum; we use Darcy's law to estimate this as $P_{v}(x)=\mu_{o}(Q/A_{c}) (L-x)/k_{c}$ \cite{layered1}. We estimate the permeability of the coarse stratum to the oil, $k_{c}$, using the Kozeny-Carman relation $k_{c}\approx\frac{\kappa}{45}\frac{\phi^{3}}{(1-\phi)^{2}}a_{c}^{2}$ \cite{amber, phil, outlet}; the relative permeability $\kappa$ quantifies the permeability reduction resulting from trapping of the wetting fluid within the crevices of the medium, as visible in Fig. 2(a). We independently measure $\kappa\approx0.16$ using a homogeneous porous medium constructed and drained in a manner similar to the experiments reported here \cite{amber, endeffect}. 

We hypothesize that the oil begins to invade the fine stratum when the flow rate is sufficiently large for the viscous pressure at the inlet, $P_{v}(0)$, to balance the capillary pressure required to invade a pore of the fine stratum $P_{\gamma, f}$. In non-dimensional form, this criterion is \begin{equation} Ca\approx Ca^{*}\equiv\frac{2a_{c}\cos\theta}{0.19 a_{f}}\frac{a_{c}}{L}\frac{\kappa}{45}\frac{\phi^{3}}{(1-\phi)^{2}}\frac{A_{c}}{A}\end{equation} We therefore expect drainage through only the coarse stratum for $Ca<Ca^{*}$; for $Ca$ above this threshold, the oil can also begin to invade the adjacent fine stratum, in agreement with our observations [Fig. 2(a-c)]. To test this prediction quantitatively, we repeat the experiments on many different stratified porous media, varying the bead sizes $a_{c}$ and $a_{f}$, medium length $L$, and the cross-sectional areas $A$ and $A_{c}$; this enables us to vary $Ca^{*}$ over one order of magnitude, $Ca^{*}\sim 10^{-5}-10^{-4}$. For all of the media tested, we observe exclusive drainage through the coarse stratum below a threshold value of $Ca$, while above this threshold, the oil also begins to invade the adjacent fine stratum, as shown by the open and filled symbols in Fig. 3, respectively. We find that the threshold for invasion into the fine stratum is given by $Ca\approx1.7Ca^{*}$ over a broad range of $Ca^{*}$ [Fig. 3, dashed line], in close agreement with our prediction [Eq. 1].
\begin{figure}
\begin{center}
\includegraphics[width=3.4in]{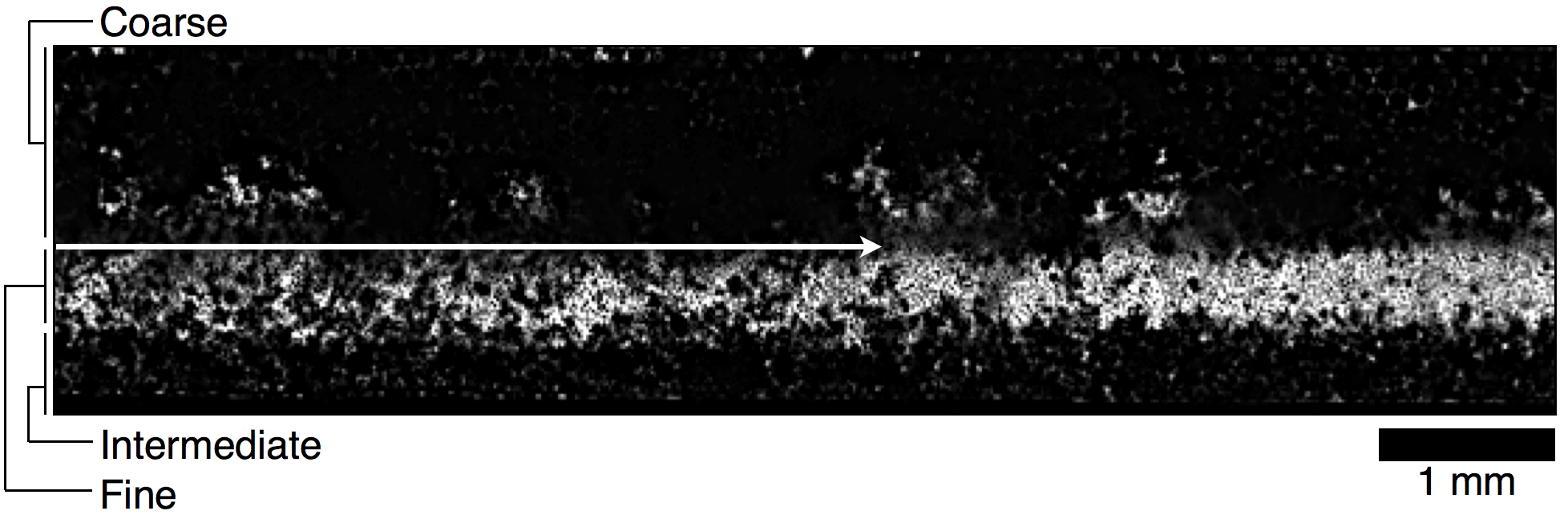}
\caption{Optical section through part of a porous medium with three strata, obtained after the flow has reached an unchanging steady state. Both coarse and intermediate strata are completely invaded by the oil (black), while the fine stratum is only partially invaded, as indicated by the arrow. Imposed flow direction is from left to right; $Ca=9.0\times10^{-5}$. The threshold $Ca^{*}=2.4\times10^{-5}$ and $5.5\times10^{-5}$ for the intermediate and fine strata, respectively.}  
\end{center}
\end{figure}

Within this picture, for sufficiently large $Ca$, oil is forced into the fine stratum not only from the inlet, but also laterally, from the adjacent coarse stratum \cite{crossflow1, crossflow2, crossflow3}. By directly visualizing the drainage dynamics at $Ca>Ca^{*}$, we confirm this lateral flow, as indicated by the arrows in Fig. 4. We therefore expect that the oil invades the fine stratum for all $x\leq L_{f}$, where $P_{v}$ exceeds $P_{\gamma, f}$ \cite{continuum}. Balancing these pressures yields \begin{equation} L_{f}/L=1-Ca^{*}/Ca \end{equation} We test this prediction by measuring the variation of $L_{f}$ with $Ca$ for the different stratified porous media, characterized by $Ca^{*}\sim10^{-5}-10^{-4}$. For all of the experiments, we find that $L_{f}/L$ increases with increasing $Ca$ [inset to Fig. 5], consistent with our expectation. Moreover, the data for different porous media collapse when $Ca$ is rescaled by $Ca^{*}$, as shown in Fig. 5, in agreement with Eq. 2. The findings presented in Fig. 3 suggest that $Ca^{*}$ should be replaced by $1.7Ca^{*}$ in Eq. 2; this yields an excellent fit to the data, as shown by the dashed line in Fig. 5. Our model thus captures both the onset and the spatial extent of oil invasion into the fine stratum.

To test the generality of our results, we also study media with three different strata: a coarse stratum, comprised of beads with radius $a_{c}=75\mu$m, a fine stratum, comprised of beads with radius $a_{f}=19\mu$m, and an intermediate stratum separating the two, comprised of beads with radius $a_{m}=38\mu$m. We observe flow behavior similar to the case of two strata: for $Ca=1.8\times10^{-5}$ and $4.8\times10^{-5}$, the oil flows through the entire coarse stratum, and also partially invades the intermediate stratum \cite{finelayer}; the spatial extent of this invasion increases with increasing $Ca$. Moreover, at an even higher $Ca=9.6\times10^{-5}$, the oil also partially invades the fine stratum. The partial invasion into the intermediate stratum requires the lateral flow of oil from the coarse stratum; we thus expect that, if the positions of the intermediate and fine strata are switched, the intermediate stratum becomes {\it completely}, not partially, invaded above a threshold $Ca$. To test this idea, we study a medium with the fine stratum separating the coarse and the intermediate strata. At a $Ca=4.5\times10^{-5}$, the oil completely invades both coarse and intermediate strata \cite{finelayer}, in contrast with the previous case, and in direct agreement with our expectation [top and bottom strata in Fig. 6]. Moreover, at an even higher $Ca=9.0\times10^{-5}$, the oil partially invades the fine stratum [arrow in Fig. 6], consistent with the picture presented here. These observations confirm that the flow path taken by the oil depends not only on the geometry of the individual strata, but also on their relative positions.

\section{Conclusions}
Using direct visualization by confocal microscopy, we demonstrate how stratification alters the path taken by a non-wetting fluid as it drains a 3D porous medium. For sufficiently small $Ca$, drainage proceeds only through the coarsest stratum of the medium; above a threshold $Ca$, the non-wetting fluid is also forced laterally, into part of the adjacent, finer strata. Our results highlight the essential role played by pore-scale capillary forces, which are frequently neglected from stratum-scale models of flow, in determining this behavior. Because geological formations are frequently stratified, we expect that our work will be relevant to a number of important applications, including understanding oil migration \cite{sorbie,chaudhari}, preventing groundwater contamination \cite{waste1,waste2}, and sub-surface storage of CO$_{2}$ \cite{bryantco2}.

\acknowledgments
It is a pleasure to acknowledge E. Amstad, D. L. Johnson, T. S. Ramakrishnan, and J. R. Rice for useful discussions, and the anonymous referees for extremely useful feedback. This work was supported by the AEC, the NSF (DMR-1006546), the Harvard MRSEC (DMR-0820484), and ConocoPhillips.


\begin{thebibliography}{10}
\bibitem{capnum}
We adopt the definition of $Ca$ used in \cite{invasion1}; however, we note that the ``local" capillary number $\mu_{o}(Q/\phi A_{c})/\gamma$ better reflects the ratio of viscous to capillary forces.
\bibitem{endeffect}
We measure the variation of the pressure drop across the medium with flow rate, before and after drainage at a single flow rate. For simplicity, we assume a constant viscous pressure gradient in the medium, and do not consider the role of end effects, or the variation of $\kappa$ with the drainage flow rate. Our measurement thus provides an order-of-magnitude estimate of the relative permeability.
\bibitem{outlet}
The viscous pressure drop in the oil as it flows through the tubing is over an order of magnitude smaller than $P_{\gamma, c}$. We thus assume that it is negligible, and take the pressure in the oil to be zero at the outlet of the porous medium.
\bibitem{interface}
We keep spatial frequencies smaller than $a_{f}/5$. Varying the cut-off frequency, as well as determining $L_{f}$ differently e.g. by identifying the position at which the intensity reaches half its maximum value, does not significantly affect the results.
\bibitem{continuum}
For simplicity, we only consider the pressures averaged laterally, over the width of each stratum \cite{yortsos, howie}.
\bibitem{finelayer}
For even smaller $Ca$, we also expect a regime, unexplored here, in which only the coarse stratum is invaded.
\end{thebibliography}
\end{document}